\documentclass[a4paper]{article}

\usepackage[english]{babel}
\usepackage[utf8x]{inputenc}
\usepackage[T1]{fontenc}

\usepackage[a4paper,top=3cm,bottom=2cm,left=3cm,right=3cm,marginparwidth=1.75cm]{geometry}

\usepackage{amsmath}
\usepackage{amssymb}
\usepackage{amsthm}
\usepackage{graphicx}
\usepackage[round]{natbib}
\usepackage{textcomp}
\usepackage{url}
\usepackage[colorinlistoftodos]{todonotes}
\usepackage[colorlinks=true, allcolors=blue]{hyperref}

\renewcommand{\vec}[1]{\mathbf{#1}}
\newcommand{\fourier}[1]{\mathcal{F}\left\{{#1}\right\}}
\newcommand{\sinc}[1]{\operatorname{sinc}\left({#1}\right)}
\newcommand{\Conv}{
  \mathop{\scalebox{1.9}{\raisebox{-0.2ex}{$\circledast$}}
  }
}

\title{A short letter on the dot product \\ between rotated Fourier transforms}
\author{Aaron R. Voelker \\
Applied Brain Research Inc.\\
Technical Report}

\begin{document}
\maketitle

\begin{abstract}
Spatial Semantic Pointers~(SSPs) have recently emerged as a powerful tool for representing and transforming continuous space, with numerous applications to cognitive modelling and deep learning.
Fundamental to SSPs is the notion of ``similarity'' between vectors representing different points in $n$-dimensional space -- typically the dot product or cosine similarity between vectors with rotated unit-length complex coefficients in the Fourier domain.
The similarity measure has previously been conjectured to be a Gaussian function of Euclidean distance. 
Contrary to this conjecture, we derive a simple trigonometric formula relating spatial displacement to similarity, and prove that, in the case where the Fourier coefficients are uniform i.i.d., the expected similarity is a product of normalized $\operatorname{sinc}$ functions: $\prod_{k=1}^{n} \sinc{a_k}$
, where $\vec{a} \in \mathbb{R}^n$ is the spatial displacement between the two $n$-dimensional points.
This establishes a direct link between space and the similarity of SSPs, which in turn helps bolster a useful mathematical framework for architecting neural networks that manipulate spatial structures.
\end{abstract}

\section*{Scalar Analysis}

Let $\fourier{\cdot}$ denote the discrete Fourier transform, and let $X \in \mathbb{R}^d$ be a vector such that all of the complex coefficients in $\fourier{X}$ are unit-length -- also known as a ``unitary'' Semantic Pointer~\citep[SP;][]{plate1995holographic, gosmann2018}.
Such vectors are fully determined by their polar angles in the Fourier domain, $\theta \in \mathbb{R}^d$, i.e., the parameters:
\begin{equation} \label{eq:theta}
    \theta = \text{Imag}\left[ \ln \fourier{X} \right] \text{,} \quad  |\theta| < \pi \text{.}
\end{equation}
Given any $x \in \mathbb{R}$, we then use the following definition to encode $x$ into a high-dimensional vector:
\begin{equation} \label{eq:fractional-binding}
    X^x \overset{\text{DEF}}{=} \mathcal{F}^{-1} \left\{ e^{i \theta x} \right\} \text{,}
\end{equation}
which combines the Spatial Semantic Pointer~(SSP) ``fractional binding'' definition from \citet{komer2019} with Euler's formula.
Essentially, $X^x$ encodes a real-valued scalar quantity ($x$) as a high-dimensional unit-length vector that may be convolved with other vectors in semantically meaningful ways, thus enabling the manipulation of topological structures within neural networks~\citep{Komer:2020, dumont2020}.

Now, consider two scalar SSPs displaced by $a \in \mathbb{R}$, as in:
\begin{equation} \label{eq:AB}
    A = X^x \text{,} \quad B = X^{x + a} \text{.}
\end{equation}
Our goal is to characterize $A^T B$, i.e., the dot product between $A$ and $B$.
Since both vectors are unitary, and the Fourier transform is unitary (i.e.,~preserves the dot product, up to a constant rescaling by $d$) and Hermitian, we can assert the following string of equalities:
\begin{equation} \label{eq:general-derive}
    d A^T B %
    = \fourier{A}^T \overline{\fourier{B}}
    = \sum_{j=1}^{d} e^{i\theta_j x - i \theta_j (x + a)}  %
    = \sum_{j=1}^{d} \text{Real}\left[ e^{ i \theta_j a } \right] %
    = \sum_{j=1}^{d} \cos \left( \theta_j a \right) \text{.}
\end{equation}
Thus, we obtain the following trigonometric formula relating the cosine similarity to the displacement $a$, in terms of the polar angles of $X$:
\begin{equation} \label{eq:general}
    A^T B = \frac{1}{d} \sum_{j=1}^{d} \cos \left( \theta_j a \right) \text{.}
\end{equation}
That is, the similarity is equal to the real-valued mean across the complex numbers that are determined by scaling each polar angle~($\theta$) by the displacement~($a$).\footnote{The imaginary components cancel out since $X$ is real, by the Hermitian symmetry of the Fourier transform.}

\begin{figure}
    \centering
    \includegraphics[width=0.6\textwidth]{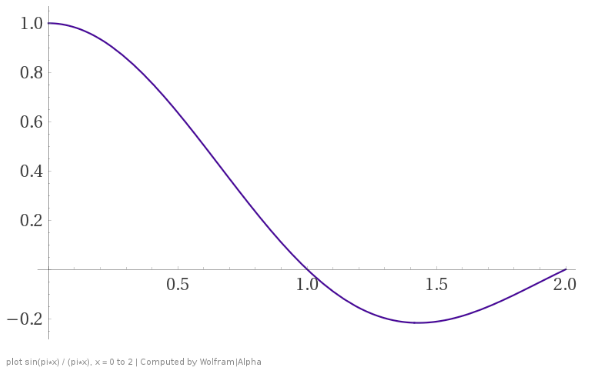}
    \caption{Plot of $\sinc{a} = \sin \left( \pi a \right) / \left( \pi a \right)$, relating the displacement ($x$-axis) to the expected similarity ($y$-axis) between two unitary SPs. The similarity is symmetric about $a=0$.
    }
    \label{fig:formula}
\end{figure}

To turn this formula into something more concrete, we must assume \emph{something} about $\theta$.
A very natural assumption is that $\theta_j \sim U\left(-\pi, \pi \right)$ are independent and identically distributed~(i.i.d.), although we note this is not the case for SSP encodings that use hexagonal lattices or other regular grids~\citep{dumont2020, Komer:2020}.
Focusing on the uniform case, we apply the law of the unconscious statistician to derive the expected similarity:
\begin{equation} \label{eq:formula}
    \mathbb{E}_\theta \left[ A^T B \right] %
    = \frac{1}{d} \sum_{j=1}^{d} \frac{1}{2 \pi} \int_{-\pi}^{\pi}  \cos \left( \theta a \right) d\theta %
    = \frac{1}{2 \pi} \int_{-\pi}^{\pi}  \cos \left( \theta a \right) d\theta %
    = \sin \left( \pi a \right) / \left( \pi a \right) \overset{\text{DEF}}{=} \sinc{a} \text{.}    \tag*{$\qed$}
\end{equation}
Here, $\sinc{\cdot}$ is defined to be the normalized $\operatorname{sinc}$ function -- plotted in Figure~\ref{fig:formula} for reference.\footnote{\url{https://www.wolframalpha.com/input/?i=plot+sin\%28pi*x\%29+\%2F+\%28pi*x\%29\%2C+x+\%3D+0+to+2}}

\section*{Higher-Dimensional Spaces}

To generalize this to SSPs representing $n$-dimensional space (e.g.,~$n=2$ in \citet{komer2019}), we repeat the above recipe, where instead $\vec{X} \in \mathbb{R}^{n, d}$ and $\Theta \in \mathbb{R}^{n, d}$ are matrices, such that equations~\ref{eq:theta} and~\ref{eq:fractional-binding} hold for each row of $\vec{X}$ and $\Theta$. 
Now, with $\vec{x}, \vec{a} \in \mathbb{R}^n$, equation~\ref{eq:AB} becomes:
\begin{equation}
    A = \Conv_{k=1}^{n} {\vec{X}_k}^{x_k} \text{,} \quad B = \Conv_{k=1}^{n} {\Vec{X}_k}^{x_k + a_k} \text{.} 
\end{equation}
Redoing equations~\ref{eq:general-derive} and~\ref{eq:general} yields:
\begin{equation}
    \boxed{ A^T B = \frac{1}{d} \sum_{j=1}^{d} \text{Real} \left[ \exp\left\{ i \sum_{k=1}^{n} \Theta_{k,j} a_k \right\} \right] = \frac{1}{d} \sum_{j=1}^{d} \cos \left( \sum_{k=1}^{n} \Theta_{k,j} a_k \right) } \,\text{.}
\end{equation}
Finally, for i.i.d. uniform $\Theta_{k, j}$, we obtain the following concrete equation for expected similarity:
\begin{equation} \label{eq:higher-d-formula}
    \mathbb{E}_\Theta \left[ A^T B \right] %
    = \frac{1}{\left(2 \pi\right)^n} \underbrace{\int_{-\pi}^{\pi} \cdots \int_{-\pi}^{\pi}}_{\text{$n$ integrals}} \cos \left( \sum_{k=1}^{n} \theta_k a_k \right) d\theta_1 \cdots d\theta_n %
    = \prod_{k=1}^{n} \sin \left( \pi a_k \right) / \left( \pi a_k \right) = \prod_{k=1}^n \sinc{a_k} \text{.}    \tag*{$\qed$}
\end{equation}


\begin{figure}
    \centering
    \includegraphics[width=\textwidth]{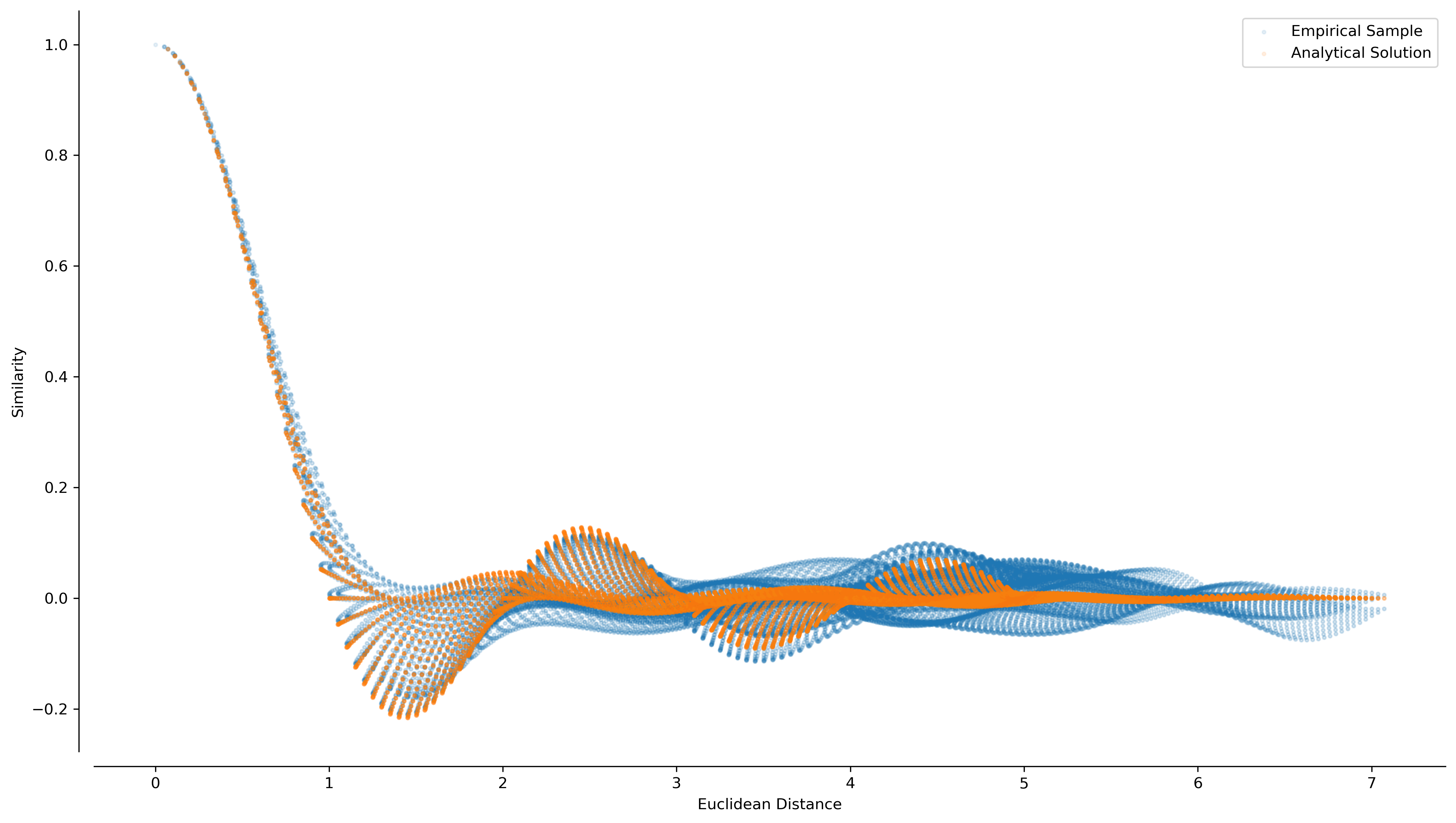}
    \caption{Empirical validation of our main result ($n = 2$,~$d = 1024$).
    Two unitary vectors~($\vec{X}$) are randomly generated with uniformly distributed polar angles~($\Theta$), and the similarity is evaluated across a square grid of displacements, $\vec{a} \in [-5, 5]^2$. 
    For each displacement, we plot the Euclidean distance ($\|\vec{a}\|$) against the actual similarity ($A^T B$) in blue (empirical) as well as the expected similarity ($\prod_{k=1}^n \sinc{a_k} = \prod_{k=1}^{n} \sin \left( \pi a_k \right) / \left( \pi a_k \right)$) in orange (analytical).
    }
    \label{fig:ssp_to_feet_analytical}
\end{figure}

\begin{figure}
    \centering
    \includegraphics[width=0.8\textwidth]{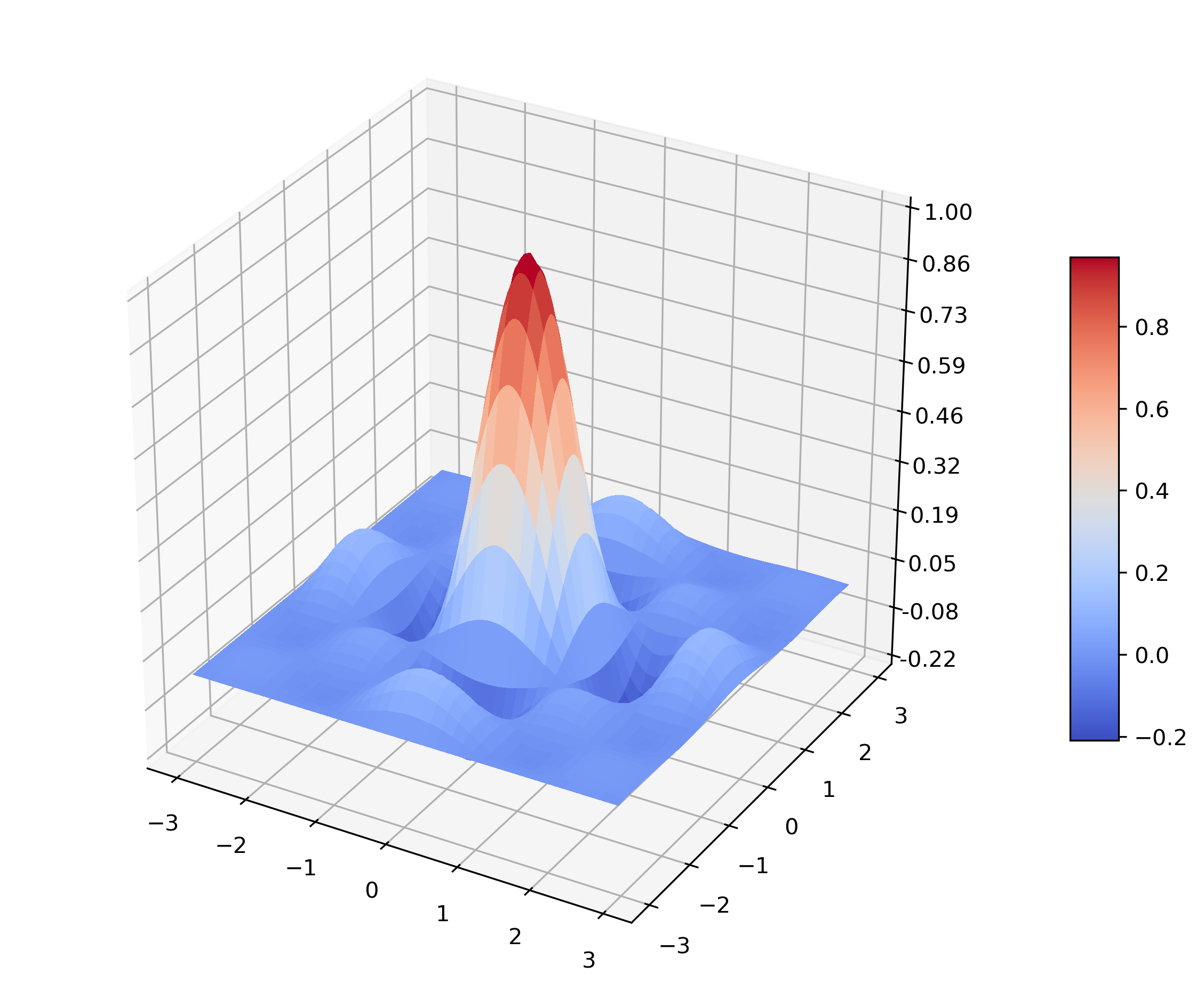}
    \caption{Surface plot of $\prod_{k=1}^n \sinc{a_k} = \prod_{k=1}^{n} \sin \left( \pi a_k \right) / \left( \pi a_k \right)$ for $\vec{a} \in [-3, 3]^2$, $n=2$, modelling the representation of an SSP encoding a two-dimensional point in space.}
    \label{fig:ssp_to_feet_surface}
\end{figure}

\newpage
\section*{Acknowledgements}

I'd like to acknowledge Ben Morcos for motivating this investigation with his questions about the observed complex structure in SSP similarity and providing the empirical sample shown in Figure~\ref{fig:ssp_to_feet_analytical}, Brent Komer for communications surrounding the distribution of SSP similarity given different choices of $X$ vectors, and Chris Eliasmith for pointing out the connection to the $\operatorname{sinc}$ function and for providing feedback.
This work was done in connection with research funded by the Laboratory for Physical Sciences~(LPS).

\bibliographystyle{plainnat}
\bibliography{refs}

\end{document}